\documentstyle[aps,preprint]{revtex}
\begin{document}
\draft

\title{
\centerline{\bf COLLAPSE OF A CIRCULAR}
\centerline{\bf LOOP OF COSMIC STRING}
}
\author{David Garfinkle}
\address{
\centerline{Department of Physics, Oakland University, Rochester, MI
48309}
\centerline{\tt garfinkl@vela.acs.oakland.edu}
}

\author{G. Comer Duncan}
\address{
\centerline{Department of Physics and Astronomy, Bowling Green
State University,
Bowling Green, OH 43403}
\centerline{\tt gcd{\rm @}einstein.bgsu.edu}
}
\maketitle

\null\vspace{-1.75mm}

\begin{abstract}
We study the collapse of a circular loop of cosmic string.  The gravitational
field of the string is treated using the weak field approximation.  The
gravitational radiation from the loop is evaluated numerically.  The metric
of the loop near the point of collapse is found analytically.
\end{abstract}

\narrowtext

\section{INTRODUCTION}

In the cosmic string scenario\cite{vil} a network of strings forms in the early
universe and evolves.  This network contains loops of various sizes and shapes.
Typically a loop oscillates between some minimum and maximum size.  The
oscillation produces gravitational radiation and the loop gradually shrinks
and disappears.  If, however as the loop oscillates its minimum size becomes
 smaller than its
Schwarzschild radius, one would expect the loop to collapse to form a
black hole with a concomitant emission of gravitational radiation.
In particular, a circular loop, whose minimum radius is zero,
would be expected to form a black hole.
Now in the case of a circular loop of cosmic string there are two effects
which could be of significant influence in the dynmaics of the
gravitational field that would be produced. We take the point of view
that up to times at which the loop radius is a few orders of magnitude
larger than the Schwarzschild radius, the weak field approximation for the
metric should be a reliable guide to the character of the fields produced.
As the loop radius monotonically decreases, its speed quickly approaches
the speed of light. Thus one would expect that regions inside the loop
would not have been affected by the data on or outside the loop for such
late times. However, there is another aspect of the dynamics in operation
here. The loop, while quickly approaching light speed, is also accelerating.
One would thus expect the world-sheet to exhibit a build-up of larger
and larger fields in the vicinity of the loop.  In the collapsing loop
problem, it would seem that both effects are in operation and neglecting
one in favor of the other would be a dubious move.
Hawking\cite{haw} has argued
that a circular loop does form a black hole and that no more than
29\% of the mass of the loop
is lost in gravitational radiation.  His argument is based on the
approximation that the loop moves at the speed of light and that the
metric of the interior of the loop is flat.


In this paper we treat the collapse of a circular loop in the
approximation of weak
field gravity.  In section II we set up the equations for the metric of the
loop.  Section III contains a numerical evaluation of the gravitational
radiation produced by the loop, including the power radiated per solid
angle and the total power radiated up to late times.  In section IV
we find an analytic expression
for the tidal force due to the loop in the limit as the loop gets small.
We show that the interior of the loop is flat. We also find  the late-time
weak field metric compatible with the tidal force for the region exterior
to but nearby a critical cylinder where the fields exhibit a large spatial
variation.
Section V contains our conclusions.  Throughout this paper we use units
in which $ c = G = 1$.

\section{METRIC OF THE LOOP}

In Minkowski spacetime the motion of a loop of string is specified by
$ {x^\mu} = {f^\mu} (\sigma , \tau ) $,
that is by giving the spacetime coordinates of the string as functions of
two worldsheet parameters
$\sigma $
and
$ \tau $.  With the usual choice\cite{vil} of the parameters
$ \sigma $
and
$ \tau $
the string equation of motion can be shown to be
$ {{\ddot f}^\mu} - {{f''}^\mu}  = 0$.
Here a dot denotes derivative with respect to
$ \tau $
and a prime is derivative with respect to
$ \sigma $.
The solution of this equation for a circular loop is
$ {f^0} = \tau $
and
\begin{equation}
{\vec f} = - \, {r_0} \, \sin \, {\tau \over {r_0}} \; \left ( {\hat x} \, \cos
\,
{\sigma \over {r_0}} \; + \; {\hat y} \, \sin \, {\sigma \over {r_0}} \right )
\; \; \; .
\end{equation}
This is a loop that begins at rest with radius
$ r_0 $
at time
$ - \pi {r_0} / 2 $
and collapses down to a radius of zero at time zero.

The stress-energy of the string is\cite{vil}
\begin{equation}
{T^{\mu \nu}} = \mu {\int _0 ^{r_0}} \; d \sigma \left ( {{\dot f}^\mu} {{\dot
f}^\nu}  \, - \, {{f '}^\mu} {{f '}^\nu} \right ) \, {\delta ^{(3)}} \left (
{\vec x} - {\vec f} (\sigma , t) \right )
\end{equation}
where
$ \mu $
is the mass per unit length of the string.

In the weak field approximation the metric is
$ {\eta _{\mu \nu}} \, + \, {h_{\mu \nu}} $
where
$ {\eta _{\mu \nu}} $
is the flat metric and
$ {h_{\mu \nu}} $
is small.  Einstein's equation is then
\begin{equation}
\left ( - \; {{\partial ^2} \over {\partial {t^2}}} \; + \; {\nabla ^2} \right
) \; { h_{\mu \nu}} = - \, 16 \pi {S_{\mu \nu}}
\end{equation}
where
$ {S_{\mu \nu}} \equiv {T_{\mu \nu}} \; - \; {1 \over 2} \; T {\eta _{\mu \nu}}
$.
In obtaining equation (3) we have also used the harmonic coordinate condition
$ {\partial ^\mu} {{\bar h}_{\mu \nu}} = 0 $
where
$ {{\bar h}_{\mu \nu}} \equiv {h_{\mu \nu}} - {1 \over 2} h {\eta _{\mu \nu}}$.
Suppose initial data for
$ h_{\mu \nu } $
is given on the surface
$ t = {t_0} $.
Then the solution of equation (3) is
$ {h_{\mu \nu}} = {H_{\mu \nu}} + {I_{\mu \nu}} $
where
$ {H_{\mu \nu}} $
is the part due to the stress-energy and
$ {I_{\mu \nu}} $
is the part due to the initial data.\cite{fut} (see Fig.\ \ref{fig1}).

First consider
$ {H_{\mu \nu}} $.
Let
$ x^\mu $
be the point at which we evaluate
$ {H_{\mu \nu}} $.
Let
$ f^\mu $
be a point on the worldsheet of the string and on the past light cone of
$ x^\mu $.
Define
$ k^\mu $
and
$ W $
by
$ {k^\mu } \equiv {x^\mu} - {f^\mu} $
and
$ W \equiv - {k_\mu} {{\dot f}^\mu} $.
Then
$ {H_{\mu \nu}} $
is given by
\begin{equation}
{H_{\mu \nu}} = 4 \mu \, {\int _{\sigma _-} ^{\sigma _+}} \; d \sigma \;
{{F_{\mu \nu }} \over W} \; \; \; .
\end{equation}
Here
$ F_{\mu \nu} $
is given by
$ {F_{\mu \nu}} \equiv  {{\dot f}_\mu} \, {{\dot f}_\nu} \; - \; {f_\mu}' \,
{f_\nu}' \; - \; {\eta _{\mu \nu }} \, {{\dot f}^\alpha} {{\dot f}_\alpha }  $
and the integral is over all points in the past light cone of
$ x^\mu $
and to the future of the initial data surface.  Equation (4) can be derived
using the Green's function for the wave equation
and the form of the string stress-energy tensor.  Define the angle
$ \lambda $
by
$ \lambda \equiv (\sigma / {r_0} ) \, - \, \phi $.
Then equation (4) becomes
\begin{equation}
{H_{\mu \nu}} = 4 \mu \, {r_0} \,  {\int _{\lambda _-} ^{\lambda _+}} \; d
\lambda \; {{F_{\mu \nu }} \over W} \; \; \; .
\end{equation}

Since
$ f^\mu $
is in the past light cone of
$ x ^\mu $
we have
$ {k^\mu } {k_\mu} = 0 $.
It then follows from equation (1) that
\begin{equation}
0 = \left ( {r^2} - {t^2} \right ) \; + \; \left ( {r_0 ^2} {\sin ^2} {\tau
\over {r_0}} \; - \; {\tau ^2} \right ) \; + \; 2 t \tau \; + \; 2 r \sin
\theta \, {r_0} \, \sin {\tau \over {r_0}} \, \cos \lambda \; \; \; .
\end{equation}
Similarly from equation (1) it follows that
$ W $
is given by
\begin{equation}
W = t \; - \; \tau \; + \; r  \; \cos {\tau \over {r_0}} \, \sin \theta
\, \cos \lambda \; + \; {r_0} \; \sin {\tau \over {r_0}} \,  \cos {\tau \over
{r_0}} \; \; \; .
\end{equation}
Equation (6) gives
$ \lambda $
as a function of
$ \tau $
or gives
$ \tau $
as an implicit function of
$ \lambda $.
This result will be used in the next section to numerically integrate equation
(5) and in section IV for an analytical treatment in the limit as the loop gets
small.

Now consider
$ {I_{\mu \nu}} $.
Let
$ S $
be the intersection of the past light cone of
$ x^\mu $
with the initial data surface.  Then
$ S $
is a sphere of radius
$ R = t - {t_0 } $.
Define
$ {P_{\mu \nu}} \left ( {\vec r} \right ) $
and
$ {V_{\mu \nu}} \left ( {\vec r} \right ) $
by
$ {P_{\mu \nu}} \left ( {\vec r} \right ) \equiv {h_{\mu \nu}} \left ( {t_0}
,{\vec r} \right ) $
and
$ {V_{\mu \nu}} \left ( {\vec r} \right ) \equiv {{\dot h}_{\mu \nu}} \left (
{t_0} ,{\vec r} \right ) $.
Then we have
\begin{equation}
{I_{\mu \nu}} = {1 \over {4 \pi R}} \; {\int _S} \; \left [
{R^{- 1}}\, {P_{\mu \nu}} \; + \; {\vec n} \cdot
{\vec \nabla} {P_{\mu \nu}} \; + \; {V_{\mu \nu}}  \right ] \; {R^2} \,
d \Omega \; \; \; .
\end{equation}
Here the natural volume element on
$ S $
is used and
$ {\vec n} $
is the outward pointing normal to
$ S $.

To find
$ I_{\mu \nu} $
we must first choose the initial data
$ P_{\mu \nu} $
and
$ V_{\mu \nu} $.
We choose the time
$ t_0 $
to be
$ - \pi {r_0} /2 $,
when the loop is at rest.  Since
$ t_0 $
is a moment of time symmetry for the loop we choose
$ {V_{\mu \nu}} = 0 $.
The initial data must be chosen consistent with the harmonic coordinate
condition
$ {\partial ^\mu} {{\bar h}_{\mu \nu}} = 0 $.
We choose
$ P_{\mu \nu } $
to be the solution of
$ {\nabla ^2} {P_{\mu \nu}} = - \, 16 \pi {S_{\mu \nu}} $.
It then follows that
$ I_{\mu \nu} $
is given by
\begin{equation}
{I_{\mu \nu}} ({\vec r} , t) \; =  \; {P_{\mu \nu}} ( {\vec r} ) \; - \; 4 \,
{\int _V} \; {d^3} r \, ' \; {{{S_{\mu \nu}} ( {\vec r} \, ' , {t_0} )} \over
{\left | {\vec r} \, - \, {\vec r} \, ' \right | }} \; \; \; .
\end{equation}
Here
$ V $
is the interior of the sphere
$ S $.

\section{GRAVITATIONAL RADIATION}

We now calculate the gravitaional radiation produced by the loop.  Define the
retarded time
$ u $
by
$ u \equiv t - r $.
Then the radiation zone is given by the limit as
$ r \to \infty $
with
$ u $
remaining finite.  All calculations in this section will be done in the
radiation zone.  The power radiated per unit solid angle in gravitational
radiation is given by\cite{mtw}
\begin{equation}
{{d P} \over {d \Omega}} = {{r^2} \over {32 \pi }} \; \left (
{{\partial {{\bar h}^{\alpha \beta }}} \over {\partial u}} \; {{\partial {{\bar
h}_{\alpha \beta }}} \over {\partial u}}\; - \; {1 \over 2} \; {{\left [
{{\partial {\bar h}} \over {\partial u}} \right ] }^2} \right ) \; \; \; .
\end{equation}
In the radiation zone the harmonic coordinate condition becomes
$ \partial {h^{\alpha u}} / \partial u = 0  $.
Since the loop is not rotating we have
$ {h_{\phi [ \alpha }} {\partial _{\beta ]} \phi } = 0 $.
It follows from these two conditions that
\begin{equation}
{{d P} \over {d \Omega}} = {1 \over {16 \pi }} \; {{\left ( {{\partial {
h_{\phi \phi}} / \partial u} \over {r {\sin ^2} \theta }} \right ) }^2} \; \;
\; .
\end{equation}
We therefore need to find
$ H_{\phi \phi} $
and
$ I_{\phi \phi} $
in order to find the power radiated.

Define
$ \psi $
and
$ w $
by
$ \psi \equiv \tau / {r_0} $
and
$ w \equiv u / {r_0} $.
The radiation zone limit of equations (6) and (7) yields
\begin{equation}
w = \psi \; + \; \sin \psi \, \sin \theta \, \cos \lambda \; \; \; ,
\end{equation}
\begin{equation}
W = r \left ( 1 \; + \; \cos \psi \, \sin \theta \, \cos \lambda \right ) \; \;
\; .
\end{equation}
{}From equation(1) we find that
$ {F_{\phi \phi }} = {r^2} {\sin ^2} \theta {\sin ^2} \lambda $.  Now define
$ B $
by
$ {B^{- 1}} \equiv 1 \; + \; \cos \psi \, \sin \theta \, \cos \lambda $.
Then we have
$ W = r {B^{ - 1}} $
and thus
\begin{equation}
{{H_{\phi \phi}} \over {r {\sin^2} \theta}} = 4 \, \mu \, {r_0} \; {\int
_{\lambda_-} ^{\lambda_+}} \; d \lambda \; B \; {\sin ^2} \lambda \; \; \; .
\end{equation}

The range of integration for
$ \lambda $
is restricted by the fact that
$ \psi > - \pi / 2 $.
Thus
$ \lambda _\pm $
are given by
\begin{equation}
{\pi \over 2} \; + \; w \; + \; \sin \theta \, \cos {\lambda _\pm} = 0 \; \; \;
{}.
\end{equation}
If
$ (\pi / 2) + w > \sin \theta $
then the range of
$ \lambda $
is from
$ 0 $
to
$ 2 \pi $.

We now find
$ I_{\phi \phi } $.
Taking the radiation zone limit of equation (9) and using equation (2) we find
\begin{equation}
{I_{\phi \phi }} = \; {P_{\phi \phi}} ( {\vec r} ) \; - \; 4 \mu {r_0} r \,
{\sin ^2} \theta \; {\int _{\lambda_-} ^{\lambda_+}} \; d \lambda \; {\sin ^2}
\lambda \; \; \; .
\end{equation}
It then follows that
\begin{equation}
{{h_{\phi \phi}} \over {r {\sin ^2} \theta}} = \; {{{P_{\phi \phi}} ( {\vec r}
)} \over {r {\sin ^2} \theta }} \; - \; 4 \mu {r_0} \; {\int _{\lambda_-}
^{\lambda_+}} \; d \lambda \; ( B \, - \, 1) \;{\sin ^2} \lambda \; \; \; .
\end{equation}

We can now compute the power radiated.  Taking the derivative of equation (17)
with respect to
$ u $
we find
\begin{equation}
{{\partial {h_{\phi \phi}} / \partial u} \over {r {\sin ^2} \theta}} = 4 \mu \;
{\int _{\lambda_-} ^{\lambda_+}} \; d \lambda \;  {{d B} \over {d w} } \;{\sin
^2} \lambda \; \; \; .
\end{equation}
(Note: though
$ \lambda _\pm $
depend on
$ u $,
the quantity
$ B - 1 $
vanishes at
$ \lambda = {\lambda _\pm } $.
Thus there is no dependence on
$ d {\lambda _\pm} / d u $
in equation (18) ).

Equation (12) gives
$ \lambda $
as a function of
$ \psi $;
however, this equation gives
$ \psi $
only as an implicit function of
$ \lambda $.
We therefore change the variable of integration from
$ \lambda $
to
$ \psi $.
Define the integral
$ K $
by
\begin{equation}
K \equiv {\int _{\psi _-} ^{\psi _+}} \; d \psi \; \left | {{d \lambda } \over
{d \psi }} \right | \; {\sin ^2} \lambda \; {{d B} \over {d w}} \; \; \; .
\end{equation}
Then since the range of
$ \lambda $
covers the range of
$ \psi $
twice we have
\begin{equation}
{{dP} \over {d \Omega}} = {4 \over \pi} \; {\mu ^2} \; {K^2} \; \; \; .
\end{equation}
Expressing all quantities in equation (19) as functions of
$ \psi $
we find
\begin{equation}
K = {\int _{\psi_-} ^{\psi_+}} \; d \psi \; {{(w - \psi ) \; {\sqrt {{\sin ^2}
\psi \, {\sin ^2} \theta \; - \; {{(w - \psi)}^2}}}} \over {{\sin ^2} \theta \;
{{\left [ \sin \psi \; + \; (w - \psi ) \cos \psi \right ] }^2}}} \; \; \; .
\end{equation}

The range of integration is fixed by two conditions: first
$ \psi $
must lie in the interval
$ [ - \pi /2, 0] $.
Second the argument of the square root in equation (21) must be positive.  We
numerically evaluate the integral as follows: for each value of
$ w $
and
$ \theta $
we find the allowed range of
$ \psi $.
We then numerically integrate to find the integral
$ K $
and use equation (20) to find the power radiated.

The results are shown in Fig.\ \ref{fig2} and \ref{fig3}.  Fig.\ref{fig2}
shows the power radiated as
a function of
$ w $
and
$ \theta $.
Fig.\ \ref{fig3}  shows the total power radiated as a function of
$ w $.
Note that most of the power is radiated near
$ w = 0 $
{\it i.e.}
when the loop is small.  This is because the loop starts out at rest and
then accelerates inward.  As
$ w $
approaches zero the speed of the loop approaches the speed of light.
The large speed and acceleration of the loop at late stages of the
collapse gives rise to a large amount of radiation.  Also note that
most of the power is radiated near
$ \theta = \pi / 2 \; i.e. $
near the plane of the loop.  As the loop shrinks the power becomes
more narrowly peaked around this plane.

\section{FIELDS NEAR THE POINT OF COLLAPSE}

The loop collapses down to a point at
$ t = 0 $.
We now find the gravitational field at spacetime points near the point of
collapse.  That is we consider values of
$ r $
and
$ t $
much smaller than
$ r_0 $.
For the weak field approximation to be valid the loop must be much larger than
its Schwarzschild radius.  That is we must have
$ | t | >> \mu {r_0} $.
For the cosmic strings considered in cosmology
$ \mu \sim {{10}^{- 6}} $.
Thus there is a range of
$ t , \; \mu {r_0} << | t | << {r_0} $
where the approximation of this section is valid.  We implement this
approximation by considering
$ t $
and
$ r $
to be of order 1 and taking the
$ {r_0} \to \infty $
limit.  In this limit we have
$ \tau / {r_0} << 1 $.
However, as we will see,
$ \tau $
can be much larger than
$ t $
or
$ r $.
So we will have to keep lowest order nonvanishing terms in
$ \tau / {r_0} $.

In the large
$ r_0 $
limit the tensor
$ I_{\mu \nu } $
vanishes.  It then follows that the metric perturbation is given by
\begin{equation}
{h_{\mu \nu}} = {{2 m} \over \pi} \;  {\int _{\lambda _-} ^{\lambda _+}} \; d
\lambda \; {{F_{\mu \nu }} \over W} \; \; \; .
\end{equation}
Here
$ m = 2 \pi {r_0} \mu $
is the mass of the loop.

We now examine the behavior of
$ W $.
Taking the large
$ r_0 $
limit of equations (6) and (7) we find
\begin{equation}
0 = \left ( {r^2} - {t^2} \right ) \; - \; {1 \over 3} \; {{\tau ^4} \over {r_0
^2}} \; + \; 2 \tau \left ( t \, + \, r \sin \theta \cos \lambda \right ) \; \;
\; ,
\end{equation}
\begin{equation}
W = t \, + \, r \sin \theta \cos \lambda \; - \; {2 \over 3} \; {{\tau ^3}
\over {r_0 ^2}} \; \; \; .
\end{equation}
Defining
$ U $
by
$ U \equiv  t \, + \, r \sin \theta \cos \lambda $
and using the last two equations to eliminate
$ \tau $
we find
\begin{equation}
{{\left ( W \, + \, 3 U \right ) }^3} \; \left ( W \, - \, U \right ) = {{16}
\over 3} \; {{{\left ( {r^2} \, - \, {t^2} \right ) }^3} \over {r_0 ^2}} \; \;
\; .
\end{equation}
Note that the right hand side of this equation is small.  Thus we have
$ W \approx U $
or
$ W \approx - 3 U $.

When
$ U $
passes through zero
$ W $
gets very small (in fact it approaches zero in the
$ {r_0} \to \infty $
limit.)  This makes
$ h_{\mu \nu } $
diverge in the limit.  However, it turns out that this divergence is gauge.
Physical quantities remain finite.  We will therefore work with the Riemann
tensor which is gauge invariant.  The Riemann tensor is given by
\begin{equation}
{R_{\mu \nu \alpha \beta}} = {\partial _\beta} {\partial _{[ \mu}} {h_{\nu ]
\alpha}} \; - \;
{\partial _\alpha} {\partial _{[ \mu}} {h_{\nu ] \beta}} \; \; \; .
\end{equation}
To find the Riemann tensor we begin by calculating
$ {\partial _z} {h_{\mu \nu}} $.
\begin{equation}
{\partial _z } {h_{\mu \nu }} = {{2 m} \over \pi} \; {\int _{\lambda _-}
^{\lambda _+}} \; d \lambda \; \left (
{{{\dot F}_{\mu \nu }} \over W} {\partial _z } \tau \; - \;
{{ F_{\mu \nu }} \over {W^2}} {\partial _z } W \right ) \; \; \; .
\end{equation}
{}From
$ {\partial _z} ( {k^\mu} {k_\mu} ) = 0 $
we find
$ {\partial _z} \tau = - z / W $.
Then from
$ {\partial _z} W = - {\partial _z} ( {k_\mu} {{\dot f}^\mu} ) $
we find
\begin{equation}
{\partial _z } W = {z \over W} \; \left (  {k_\mu} {{\ddot f}^\mu}
\, - \, {{\dot f}^\mu} {{\dot f}_\mu}  \right ) \; \; \; .
\end{equation}
Evaluating the quantity in parentheses in equation (28) we obtain
\begin{equation}
{\partial _z } {h_{\mu \nu }} = {{- 2 m} \over \pi} \; {\int _{\lambda _-}
^{\lambda _+}} \; {{d \lambda} \over W} \; \left (
{{{\dot F}_{\mu \nu }} \over W} \; + \;
{{ F_{\mu \nu }} \over {W^2}} 2 \; {{\tau ^2} \over {r_0 ^2}} \; \left ( 1 \; +
\; {{r \sin \theta \cos \lambda } \over { 2 \tau }} \right )  \right ) \; \; \;
{}.
\end{equation}

We also need to compute
$ {h_{z z}} $.
In the large
$ r_0 $
limit
$ {F_{z z}} = {\tau ^2} / {r_0 ^2} $
so it follows that
\begin{equation}
{h_{z z}} = {{ 2 m} \over \pi} \; {\int _{\lambda _-} ^{\lambda _+}} \; {{d
\lambda} \over W} \; {{\tau ^2} \over  {r_0 ^2}} \; \; \; \; .
\end{equation}

Define the critical cylinder to be the surface where
$ r \sin \theta = - t $.
Then the behavior of the metric inside the critical cylinder is very
different from the behavior outside.

\subsection{Inside the Critical Cylinder}

We now consider the case
$ r \sin \theta < - t $,
that is inside the critical cylinder.
Then we have
$ U < 0 $
and
$ U $
is bounded away from zero.  Therefore
$ W $
is bounded away from zero.  It then follows that in the large
$ {r_0} $
limit we have
$ {h_{z z}} \to 0 $.
Furthermore since
$ {{\dot F}_{\mu \nu}} \to 0 $
in this limit it follows that
$ {\partial _z} {h_{\mu \nu}} \to 0 $.
It then follows that
$ {R_{abcz}} = 0 $.
However, the spacetime off the string is vacuum.  So the Riemann
tensor is equal to the Weyl tensor and so
$ {C_{abcz}} = 0 $.
But it follows from the properties of the Weyl tensor that if
$ {C_{abcz}} = 0 $
then the Weyl tensor is zero.  So in our case the Riemann tensor
vanishes inside the critical cylinder and
the spacetime in this region is flat.

\subsection{Outside the Critical Cylinder}

Now we treat the case where
$ r \sin \theta > - t $,
that is outside the critical cylinder.
Define the variable
$ X $
by
\begin{equation}
X \equiv {{{\sqrt {r_0}} \, ( W \, + \, 3 U ) } \over {2 {{\left ( {r^2} -
{t^2} \right ) }^{3/4}}}}
\; \; \; \; .
\end{equation}
Expressing
$ W , \; U $
and
$ \tau $
in terms of
$ X $
we have
\begin{equation}
W = { {{\left ( {r^2} - {t^2} \right ) }^{3/4}} \over {2 {\sqrt {r_0}} }} \;
{{1 \, + \, {X^4}} \over {X^3}}  \; \; \; ,
\end{equation}
\begin{equation}
U = { {{\left ( {r^2} - {t^2} \right ) }^{3/4}} \over {2 {\sqrt {r_0}} }} \;
\left ( X \; - \; {1 \over 3} \; {X^{- 3}} \right )  \; \; \; ,
\end{equation}
\begin{equation}
{{\tau ^2} \over {r_0 ^2}} = {{\sqrt {{r^2} - {t^2}}} \over {{r_0} {X^2}}} \;
\; \; .
\end{equation}
{}From these equations we can express all quantities in terms of
$ X $
and
$ r_0 $.

We now compute the metric component
$ h_{z z} $.
The quantity
$ {F_{zz}} / W $
is a bounded function of
$ X $
multiplied by
$ {r_0 ^{- 1 / 2}} $.
It then follows that in the large
$ r_0 $
limit
$ {h_{zz}} \to 0 $.

Now we compute
$ {\partial _z} {h_{\mu \nu}} $.
In the large
$ r_0 $
limit the only contribution to the integral comes from the region where
$ W $
is very small.  That is the region where
$ U \approx 0 $.
This is the region where
$ \lambda \approx {\lambda _0 } $
where
$ \lambda _0 $
is given by
$ t + r \sin \theta \cos {\lambda _0} = 0 $.
In the integral for
$ {\partial _z} {h_{\mu \nu}} $
we change the variable of integration from
$ \lambda $
to
$ X $.
Since
$ \lambda \approx {\lambda _0} $
in the region of interest we should be able to evaluate all quantities at
$ \lambda = {\lambda _0} $.
Define the coordinates
$ \rho $
and
$ s $
by
$ \rho \equiv r \sin \theta $
and
$ s \equiv {\sqrt {{\rho ^2} - {t^2}}} $.
Then from equations (32) and (33) it follows that
$ d U / W = d X / X $
and therefore that
$ d \lambda / W = d X / ( s X ) $.
Our approximation that
$ \lambda \approx {\lambda _0 } $
is valid provided that the fractional change in
$ X $
is much larger than the fractional change in
$ \sin \lambda $
in the region where
$ U \approx 0 $.
That is we require
$ | d X / X | >> |d \sin \lambda / \sin \lambda | $.
This is equivalent to
$ {s^2} >> | t | {{\left ( {r^2} - {t^2} \right ) }^{3/4}} / {\sqrt {r_0}}$.
This will be satisfied except for extremely small values of
$ s $.

Now the range of
$ \lambda $
covers the range of
$ X $
twice.  So for functions that are odd in
$ \lambda $
the integral vanishes.  For functions that are even in
$ \lambda $
there is a factor of two multiplying the integral.  We then find that
$ {\partial _z} {h_{\mu \nu}} $
is given by
\begin{equation}
{\partial _z} {h_{\mu \nu}} =  { {- 4 m z } \over {\pi s}} \; {\int _{X_-}
^{X_+}} \; {{d X} \over X} \;
\left ( {{{\dot F}_{\mu \nu }} \over W} \; + \;
{{ F_{\mu \nu }} \over {W^2}} 2 \; {{\tau ^2} \over {r_0 ^2}} \; \left ( 1 \;
- \; {t \over { 2 \tau }} \right )  \right ) \; \; \; .
\end{equation}
Now express all quantities in the integrand in terms of
$ X $
and
$ r_0 $.
Then we find that
$ F_{\mu \nu } $
is of order
$ 1 $,
$ {{\dot F}_{\mu \nu}} / W $
is of order
$ r_0 ^{- 1} $,
$ \tau / (W {r_0} ) $
is of order
$ 1 $
and
$ t / \tau $
is of order
$ 1 / {\sqrt {r_0}} $.
So the
$ {{\dot F}_{\mu \nu}} $
and the
$ t / 2 \tau $
terms in the integrand give zero contributions in the large
$ r_0 $
limit. Define
$ {\bar F}_{\mu \nu} $
by
$ {{\bar F}_{\mu \nu}} \equiv  {{\left . {F_{\mu \nu}} \right |}
_{\lambda = {\lambda _0}}} $.
Then since the integrand is peaked around
$ \lambda _0 $
and since
$ F_{\mu \nu } $
is slowly varying we have
\begin{equation}
{\partial _z} {h_{\mu \nu}} =  { {- 8 m } \over \pi} \; {{ z {{\bar F}_{\mu
\nu}}} \over { s \, \left ( {s^2} + {z^2} \right ) }} \; {\int _{X_-} ^{X_+}}
\; {{4 \, {X^3} \, d X} \over  {{\left ( 1 \, + \, {X^4} \right ) }^2}} \; \;
\; .
\end{equation}
In the large
$ r_0 $
limit the limits of integration are zero and infinity.  So we find
\begin{equation}
{\partial _z} {h_{\mu \nu}} =  { {- 8 m } \over \pi} \;
{{ z {{\bar F}_{\mu \nu}}} \over {s \left ( {s^2} + {z^2} \right ) }}
\; \; \; .
\end{equation}

Evaluating
$ F_{\mu \nu} $
at
$ \lambda = {\lambda _0} $
we find
\begin{equation}
{{\bar F}_{\mu \nu}} = {\rho ^2} \, {\partial _{\mu} (t / \rho ) \, {\partial
_{\nu}} (t/\rho ) \; + \; {s^2} \, {\partial _{\mu}} \phi \, {\partial _{\nu}}
\phi
\; \; \; .
}
\end{equation}

We now compute the Riemann tensor.  Since
$ {h_{\mu z}} = 0 $
it follows that
\begin{equation}
{R_{\mu \nu \tau z}} = {1 \over 2} \; \left ( {\partial _{\mu}} {\partial _z}
{h_{\nu \tau}} \; - \; {\partial _{\nu}} {\partial _z} {h_{\mu \tau}} \right )
\; \; \; .
\end{equation}
So
$ R_{\mu \nu \tau z} $
can be computed using the expression in equation (37).  However, in vacuum
the Riemann tensor is equal to the Weyl tensor and is determined by
$ R_{\mu \nu \tau z} $.
Let
$ {\hat \phi}^{\mu} $
and
$ {\hat z}^{\mu} $
be the unit vectors in the
$ \phi $
and
$ z$
directions respectively.  Define the tensor
$ E_{\mu \nu} $
by
$ {E_{\mu \nu}} \equiv {C_{\mu {\hat \phi} \nu {\hat \phi} }} $.
Because the axial Killing vector
$ \partial / \partial \phi $
is hypersurface orthogonal, the Weyl tensor is
determined by
$ E_{\mu \nu} $.
We have
\begin{equation}
{C_{\mu \nu \tau \sigma}} = - 4 {{\hat \phi}_{[\mu}} {E_{\nu][\tau}} {{\hat
\phi}_{\sigma]}} \; - \;
{{\epsilon _{\mu \nu}}^{mp}} {{\epsilon _{\tau \sigma}}^{nq}} {E_{mn}} {{\hat
\phi}_p} {{\hat \phi}_q} \; \; \; .
\end{equation}
The tensor
$ E_{\mu \nu} $
is determined by
$ C_{\mu \nu \tau z} $
by
\begin{equation}
{E_{\mu \nu}} = \left ( {{\hat \phi}_{\mu}} {{\hat \phi}_{\nu}} -
{\eta _{\mu \nu}} \right ) \,
{C_{{\hat \phi} z {\hat \phi} z}} \; - \; {C_{\mu z \nu z}} \; - \; 2 {{\hat
z}_{(\mu}}
{C_{\nu ) {\hat \phi } {\hat \phi } z}} \; \; \; .
\end{equation}
Finally some straightforward but tedious algebra yields the following
expression for
$ E_{\mu \nu} $.
\begin{equation}
{E_{\mu \nu}} = {{4 m} \over \pi} \; {s \over {{\rho ^2} \, {{\left ( {s^2} +
{z^2} \right ) }^2}}} \; \left [ \left ( {s^2} - {z^2} \right ) \, \left (
{\partial _{\mu}} z {\partial _{\nu}} z \; - \; {\partial _{\mu}} s {\partial
_{\nu}} s \right ) \; - \;
4 z s  \; {\partial _{(\mu}} z {\partial _{\nu)}} s \right ] \; \; \; .
\end{equation}
This expression is actually singular as
$ s \to 0 $.
(that is as
$ \rho \to - t $).
This is because
$ {\partial _{\mu}} s $
is singular as
$ s \to 0 $.
($ {\partial _\rho} s = \rho / s $).
So the Cartesian components of the Riemann tensor diverge like
$ 1 / {\sqrt { \rho + t }} $
as
$ \rho \to - t $.
Fig.\ \ref{fig4} shows the Riemann tensor component
$ R_{\rho {\hat \phi } \rho {\hat \phi }} $
as a function of
$ \rho $
and $ z $.
The component $ R_{\rho {\hat \phi} \rho {\hat \phi }} $
is measured in units of $4 m / \pi {t^3} $
and the coordinates $ \rho $
and $ z $ are measured in units of $ | t |$.

It turns out that the quadratic curvature scalars are not singular (except at
$ z = s = 0$).  We have
\begin{equation}
{R^{\mu \nu \tau \sigma}} {R_{\mu \nu \tau \sigma}} = 4 {E^{\mu \nu}} {E_{\mu
\nu}} = 2 {{\left [ {{8 m} \over \pi} \; {s \over {{\rho ^2} \, \left ( {s^2} +
{z^2} \right ) }} \; \right ] }^2} \; \; \; .
\end{equation}

Note that things are becoming singular just where our approximation breaks down
({\it i.e.} for extremely small
$ \rho + t $).
However,  for moderately small
$ \rho + t $
our approximation is still good.  So we find that the tidal force shows rapid
variation and large values near the critical value of
$ \rho $.

The metric perturbation
$ h_{\mu \nu} $
diverges at late times.  However, except for the critical cylinder this
bad behavior is pure gauge.  This means that there is a well behaved
metric perturbation that gives the same Riemann tensor as
$ h_{\mu \nu} $.
Define the perturbation
$ {\tilde h}_{\mu \nu} $
by
\begin{equation}
{{\tilde h}_{\mu \nu}} \equiv - \; {{4 m} \over \pi } \; {{\bar F}_{\mu \nu}}
\;
{s^{- 1}} \, \ln \left ( {s^2} + {z^2} \right ) \; \; \; .
\end{equation}
Then it follows from equation (37) that
$ {\partial _z} {{\tilde h}_{\mu \nu}} = {\partial _z} {h_{\mu \nu}} $.
It then follows from equation (38) that
$ {\tilde h} = 0 $
and
$ {\partial ^{\mu} } {{\tilde h}_{\mu \nu}} = 0 $
so
$ {\tilde h}_{\mu \nu} $
is in harmonic gauge.  Finally some straightforward but tedious algebra
using equation (38) shows that
$ {\partial ^{\mu}} {\partial _{\mu} } {{\tilde h}_{\tau \sigma}} = 0$.
Therefore
$ {\tilde h}_{\mu \nu} $
is a solution of the linearized Einstein equations.  However, since
$ {\partial _z} {{\tilde h}_{\mu \nu}} = {\partial _z} {h_{\mu \nu}} $
it then follows that the Riemann tensor of
$ {\tilde h}_{\mu \nu} $
is the same as that of
$ h_{\mu \nu}$.
Therefore
$ {\tilde h}_{\mu \nu } $
can be regarded as the metric perturbation of the string loop
at late times exterior to the critical cylinder.

\section{CONCLUSION}

The calculations of the last two sections give a picture of the fields produced
by the collapse of a circular loop of cosmic string.  When the loop begins to
collapse it is moving slowly and its gravitational field is quite mild.  Very
little power is radiated in gravitational radiation in this early stage of
collapse.  As the loop gets smaller it moves faster and more power is radiated.
 Most of this power is radiated in directions near the plane of the loop.

When the loop gets very small its speed approaches the velocity of light.
Near the loop the tidal forces become large.  These tidal forces have a
strong spatial dependence.
We have shown that inside the critical
cylinder the tidal force vanishes so the metric is flat.  This flatness of
the metric was assumed by Hawking\cite{haw} in his argument that the loop must
form a black hole.  Thus we have supplied a missing step in Hawking's argument.

Outside the critical cylinder there is tidal force and the tidal force
becomes large as one approaches the critical cylinder (strictly speaking
calculation gives an infinite tidal force on the critical cylinder.
However
our approximation breaks down when we get too close to the
critical cylinder.  Thus all we can say is that near the critical cylinder
the tidal force has large values and large spatial variation.)

When the size of the loop gets to
$ \sim 1 / \mu $
of its original size it has approached its Schwarzschild radius.  In
this regime the weak field approximation, which we have used throughout
this paper,
breaks down.  Thus our results do not tell us what happens when the
loop collapses to form a black hole.  To treat this strong field regime
one would need to use general relativity rather than the weak field
approximation.  The
equations then become much more complicated, and there is little hope of an
analytic treatment.  A numerical treatment may be feasible.  The
results of section IV could be used as initial data for a general relativistic
numerical evolution of the string and its gravitational field.  Note, however,
that the fields have large spatial variation.  Thus one would need a computer
code that could handle large spatial changes in order to evolve the loop
until it
collapses completely to form a black hole.

\newpage

\begin{figure}
\caption{Spacetime diagram showing the intersection of the past lightcone
of the field point and the world-sheet of the collapsing string loop.}
\label{fig1}
\end{figure}

\begin{figure}
\caption{Plot of power radiated per unit solid angle as a function of
w and $ \theta $.}
\label{fig2}
\end{figure}
\begin{figure}
\caption{Plot of the total power emitted as a function of w.}
\label{fig3}
\end{figure}
\begin{figure}
\caption{Surface plot of $ E_{\rho \rho} $ as a function of $ z $ and
$ \rho $. $ E_{\rho \rho} $ is measured in units of $ 4 m / \pi {t^3}   $ while
$ \rho $ and $ z $
are measured in units of $ | t | $. }
\label{fig4}
\end{figure}
%
%
%
%
%
%
%

\end{document}